\global\def\draftcontrol{0}

   \def\versionno{ gl}

\catcode`\@=11

\expandafter\ifx\csname draftcontrol\endcsname\relax\global\def\draftcontrol{0}
\fi

{\count255=\time\divide\count255 by 60
\xdef\hourmin{\number\count255}
\multiply\count255 by-60\advance\count255 by\time
\xdef\hourmin{\hourmin:\ifnum\count255<10 0\fi\the\count255}}
\def\draftdate{\number\month/\number\day/\number\year\ \ \ \hourmin }

\newcommand\makepapertitle{\par
  \begingroup
    \renewcommand\thefootnote{\@fnsymbol\c@footnote}%
    \def\@makefnmark{\rlap{\@textsuperscript{\normalfont\@thefnmark}}}%
    \long\def\@makefntext##1{\parindent 1em\noindent
            \hb@xt@1.8em{%
                \hss\@textsuperscript{\normalfont\@thefnmark}}##1}%
     \newpage
     \global\@topnum\z@   
     \@makepapertitle
     \thispagestyle{empty}\@thanks
  \endgroup
  \setcounter{footnote}{0}%
  \global\let\thanks\relax
  \global\let\makepapertitle\relax
  \global\let\@makepapertitle\relax
  \global\let\@thanks\@empty
  \global\let\@author\@empty
  \global\let\@date\@empty
  \global\let\@title\@empty
  \global\let\title\relax
  \global\let\author\relax
  \global\let\date\relax
  \global\let\and\relax
  \def\version{\let\version\@version\@gobble}
}
\def\@makepapertitle{%
  \newpage
   \ifnum\draftcontrol=1 {}
   \version\versionno
   \vskip 3em%
   \else
   \hfill\hbox to 3cm {\parbox{4cm}{\@pubnum}\hss}%
   \vskip 3em%
   \fi
   \begin{center}%
   \let \footnote \thanks
     {\LARGE {\@title}}%
     \vskip 1.5em%
     {\normalsize
       \lineskip .5em%
       \begin{tabular}[t]{c}%
         \@author
       \end{tabular}\par}%
     \vskip 1.5em%
     {\@bstract}%
     \end{center}%
     \vskip 1.5em
     \@date%
   \par
}

\gdef\@pubnum{}
\def\pubnum#1{%
  \gdef\@pubnum{#1}}

\gdef\@bstract{}
\def\Abstract#1{%
  \gdef\@bstract{%
   \parbox{\textwidth-0pc}{%
   \centerline{\bf Abstract}\penalty1000%
\kern.2cm%
\noindent
\renewcommand\baselinestretch{1.0}%
{#1}}}
}

\def\ps@paper{\let\@mkboth\@gobbletwo%
     \ifnum\draftcontrol=1
    \def\@oddfoot{\hbox to \textwidth{\tiny \versionno \hfil\tiny\draftdate}%
    \hskip -\textwidth \hbox to \textwidth{\hfil\rm\thepage\hfil}}%
     \else\def\@oddfoot{\hbox to \textwidth{\hfil\rm\thepage\hfil}}
     \fi
     \let\@evenfoot\@oddfoot
}

\def\body{\clearpage
          \pagestyle{paper}
    }

\def\@version#1{\ifnum\draftcontrol=1
\typeout{}\typeout{#1}\typeout{}
\vskip3mm\centerline{\hbox{\fbox{\normalsize{\tt DRAFT -- #1 -- }
                   {\draftdate}}}}\vskip3mm
\fi}
\let\version\@version
\long\def\eqlabel#1{\ifnum\draftcontrol=1
                    \tag@false  
                    \tag*{(\theequation) \hbox to -0.2cm{\hspace{0cm}\small{#1}\hss}}
                    \refstepcounter{equation}
                    \edef\@currentlabel{\theequation}
                    \ltx@label{#1}          
                    \else
                    \label{#1}
                    \fi
                    }
\let\st@bibitem\@bibitem
\let\st@lbibitem\@lbibitem
\ifnum\draftcontrol=1
  \def\@bibitem#1{%
    \st@bibitem{#1}\a@@label{#1}\ignorespaces}
  \def\@lbibitem[#1]#2{%
    \st@lbibitem[#1]{#2}\a@@label{#2}\ignorespaces}
  \def\a@@label#1{%
    \gdef\a@lab{\smash{\normalfont\small#1}}
    \ifvmode
      \if@inlabel
        \global\setbox\@labels\hbox{%
          \llap{\a@lab\let\a@lab\relax
                \kern\@totalleftmargin\kern\marginparsep}%
          \box\@labels}%
      \fi
    \fi}
\fi

\documentclass[12pt,letterpaper]{article}

\usepackage{amsmath,amssymb,array,calc,epsfig,rotating,bm}
\usepackage[sort]{cite}
\usepackage{graphicx}
\usepackage{psfrag,verbatim}

\ifnum\draftcontrol=1
\fi

\renewcommand\baselinestretch{1.25}
\renewcommand\section{\@startsection {section}{1}{\z@}%
                                   {-3.5ex \@plus -1ex \@minus -.2ex}%
                                   {2.3ex \@plus.2ex}%
                                   {\normalfont\large\bfseries}}
\renewcommand\subsection{\@startsection{subsection}{2}{\z@}%
                                   {-3.25ex\@plus -1ex \@minus -.2ex}%
                                   {1.5ex \@plus .2ex}%
                                   {\normalfont\normalsize\bfseries}}
\renewcommand\subsubsection{\@startsection{subsubsection}{3}{\z@}%
                                   {-3.25ex\@plus -1ex \@minus -.2ex}%
                                   {1.5ex \@plus .2ex}%
                                   {\normalfont\normalsize\it}}
\renewcommand\paragraph{\@startsection{paragraph}{4}{\z@}%
                                   {-3.25ex\@plus -1ex \@minus -.2ex}%
                                   {1.5ex \@plus .2ex}%
                                   {\normalfont\normalsize\bf}}

\numberwithin{equation}{section}

\def\revise#1       {\raisebox{-0em}{\rule{3pt}{1em}}%
                     \marginpar{\raisebox{.5em}{\vrule width3pt\
                     \vrule width0pt height 0pt depth0.5em
                     \hbox to 0cm{\hspace{0cm}{%
                     \parbox[t]{4em}{\raggedright\footnotesize{#1}}}\hss}}}}

\newcommand\nxt[1]  {\\\fnxt#1}

\def\caln         {{\cal N}}
\def\calo         {{\cal O}}

\def\del          {\partial}

\def\Re           {{\rm Re\hskip0.1em}}
\def\Im           {{\rm Im\hskip0.1em}}

\def\sqr#1#2{{\vcenter{\vbox{\hrule height.#2pt
 \hbox{\vrule width.#2pt height#1pt \kern#1pt
 \vrule width.#2pt}\hrule height.#2pt}}}}

\def\a{\alpha}
\def\b{\beta}
\def\w{\omega}
\def\r{\rho}
\def\dd{\delta}
\def\e{\epsilon}

\def\g{\gamma}

\def\aa1{\phi}
\def\cc1{\psi}

\def\Om{\Omega}

\catcode`\@=12

\begin{document}


\title{\bf Small black holes in $AdS_5\times S^5$}

\date{April 6, 2015}

\author{
Alex Buchel$ ^{1,2}$ and Luis Lehner$ ^2$\\[0.4cm]
\it $ ^1$Department of Applied Mathematics\\
\it University of Western Ontario\\
\it London, Ontario N6A 5B7, Canada\\
\it $ ^2$Perimeter Institute for Theoretical Physics\\
\it Waterloo, Ontario N2J 2W9, Canada
}

\Abstract{We consider small black holes in $AdS_5\times S^5$, 
smeared on $S^5$. We compute the spectrum of $\ell \in [1,10]$ $S^5$-quasinormal modes 
corresponding to fluctuations leading to localization of these 
black holes on $S^5$. We recover the zero mode 
found by Hubeny and Rangamani (HR) previously \cite{Hubeny:2002xn},
and explicitly demonstrate that a  Gregory-Laflamme type instability is at play
in this system. The instability is associated with the 
expectation value of a dimension-5 operator.
}

\makepapertitle

\body

\version\versionno
\tableofcontents

\section{Introduction}\label{intro}

Small black holes in type IIb supergravity in asymptotically global $AdS_5\times S^5$ geometry 
are important in the holographic correspondence~\cite{m1,Aharony:1999ti} between string theory and gauge theories 
in four space-times dimensions (see \cite{Buchel:2015lla} for a recent discussion). 
The simplest black holes have $SO(6)$ symmetry\footnote{See \cite{Dias:2015pda} for interesting 
generalizations.} and are described by the following line element\footnote{There is also a nontrivial 
5-form flux.} 
\begin{equation}
ds_{10}^2=-f dt^2 +\frac{dr^2}{f}+r^2 (d\Om_3)^2+L^2 (d\Om_5)^2\,,\qquad f=1+\frac{r^2}{L^2}
-\frac{r_+^2}{r^2}\left(\frac{r_+^2}{L^2}+1\right)\,,
\eqlabel{bhmetric}
\end{equation}
where $L$ is the radius of the $S^5$ with the round metric $(d\Om_5)^2$,  $(d\Om_3)^2$ is the round metric 
on the $S^3$ and $r_+$ denotes the location of the regular Schwarzschild horizon. It was proposed in 
\cite{Banks:1998dd,Horowitz:1999uv} that as the black hole becomes sufficiently small,
in the limit $\frac {r_+}{L}\to 0$, it would suffer a Gregory-Laflamme (GL) instability  \cite{Gregory:1993vy},
resulting in its localization on $S^5$. The first analysis of the GL instability in this context was
performed in \cite{Hubeny:2002xn} (HR) where the authors identified, in particular, an $\ell=1$ zero mode of $S^5$
which was assumed to 
be the GL mode at the threshold of instability. This HR zero mode becomes normalizable when\footnote{A more precise 
value was computed in \cite{Dias:2015pda}. Our computation \eqref{hrres}
agrees with the  one reported in \cite{Dias:2015pda} to an accuracy of $\sim 7\times  10^{-6}$.} 
\cite{Hubeny:2002xn}
\begin{equation}
\frac{r_+}{L}\simeq 0.4259\,.
\eqlabel{glmode}
\end{equation} 

We here revisit the analysis of \cite{Hubeny:2002xn} and compute the spectrum of the $\ell=1$ quasinormal modes 
about $SO(6)$ symmetric black holes in  $AdS_5\times S^5$. We explicitly exhibit the GL instability,
and determine its onset. We find that the instability arises as foreseen 
in \cite{Banks:1998dd,Horowitz:1999uv}, and appears precisely when the HR zero mode becomes normalizable. 

In section \ref{not} we set up our notation, and discuss the 
$SO(4)\times SO(5)$ symmetric backgrounds in type IIb SUGRA. In particular, in section \ref{lisone}
we present the equations of motion describing $\ell=1$ quasinormal modes of $SO(6)$ symmetric black holes.
In section \ref{hr} we reproduce the HR zero mode. In section \ref{gl} we compute the spectrum of $\ell=1$
quasinormal modes, which are associated with the expectation value of a dimension-5 operator of a boundary SYM.
In section \ref{higher} we compute the spectrum of higher-$\ell$ quasinormal modes and demonstrate 
that they become unstable 
for smaller values of $\frac {r_+}{L}$ than \eqref{glmode}.
We conclude in section \ref{con}.

\section{$SO(4)\times SO(5)$ symmetric ansatz in type IIb SUGRA}\label{not}

The type IIb supergravity equations of motion, where only the metric $g_{\mu\nu}$ and the Ramond-Ramond five-form
$F_{(5)}$ are turned on, take the form:
\begin{equation}
G_{\mu\nu}\equiv R_{\mu\nu}-\frac{1}{48}F_{(5)\mu\a\b\g\dd}F_{(5)\nu}^{\quad \a\b\g\dd}=0\,,\qquad dF_{(5)}=0\,,\qquad F_{(5)}=\star F_{(5)}\,.
\eqlabel{2beoms}
\end{equation}
We are particularly interested in the most general ansatz describing solutions with $SO(4)\times SO(5)$ isometry. To obtain
an explicit expression for the equations determining such solutions we first
fix the reparametrization invariance such that 
\begin{equation}
g_{tx}=g_{t\theta}=g_{x\theta}=0\,.
\eqlabel{diffeo}
\end{equation}
We can thus write the line element as, 
\begin{equation}
\begin{split}
&ds_{10}^2=-c_1^2\ (dt)^2+c_2^2\ (dx)^2+c_3^2\ (d\Om_3)^2 +c_4^2\ (d\theta)^2 +c_5^2\ (d\Om_4)^2\,,\cr
&F_{(5)}=(a_0\ d\theta+a_1\ dt+a_2\ dx)\wedge d\Om_4+(a_3\ d\theta\wedge dt+a_4\ d\theta\wedge dx+a_5\ dt\wedge dx)\wedge d\Om_3\,,\\
&c_i=c_i(t,x,\theta)\,,\qquad a_i=a_i(t,x,\theta)\,,
\end{split}
\eqlabel{ansatz}
\end{equation}
where $d\Om_3$ is a volume form on a unit radius $S^3$ and  $d\Om_4$ is a volume form on a unit radius $S^4$.
Next we can eliminate $\{a_0,a_1,a_2\}$ by imposing 5-form duality,
\begin{equation}
\begin{split}
a_0=&-\frac{c_4 c_5^4\sin^4\theta}{c_1c_2c_3^3}\ a_5\,,\qquad a_1=-\frac{c_1 c_5^4\sin^4\theta}{c_2c_4c_3^3}\ a_4\,,
\qquad a_2=-\frac{c_2 c_5^4\sin^4\theta}{c_1c_4c_3^3}\ a_3\,.
\end{split}
\eqlabel{a0a1a2}
\end{equation}
The resulting equations constitute a system of partial differential equations
resulting from the eight non-trivial Einstein equations (the particular
expressions are rather involved and we will not present their explicit form at
this point):
\begin{equation}
\begin{split}
&G_{tt}=G_{xx}=G_{\Om_3\Om_3}=G_{\theta\theta}=G_{\Om_4\Om_4}=0\,,\qquad G_{tx}=G_{t\theta}=G_{x\theta}=0\,,
\end{split}
\eqlabel{eoms}
\end{equation}
together with the five-form Maxwell equations:
\begin{equation}
\begin{split}
0=&\del_\theta a_3-\frac{c_4^2}{c_2^2} \del_x a_5+4\cot(\theta) a_3
+\frac{c_4^2 a_5}{c_2^2}\del_x \ln\frac{c_1c_2c_3^3}{c_4c_5^4}
-a_3\del_\theta\ln\frac{c_1c_4c_3^3}{c_2c_5^4}\,,\\
0=&\del_\theta a_4-\frac{c_4^2}{c_1^2} \del_t a_5+4\cot(\theta) a_4
+\frac{c_4^2 a_5}{c_2^2}\del_t \ln\frac{c_1c_2c_3^3}{c_4c_5^4}
-a_4\del_\theta\ln\frac{c_2c_4c_3^3}{c_1c_5^4}\,,\\
0=&\del_t a_3-\frac{c_1^2}{c_2^2} \del_x a_4
-a_3\del_t \ln\frac{c_1c_4c_3^3}{c_2c_5^4}
-\frac{c_1^2 a_4}{c_2^2}\del_\theta\ln\frac{c_1c_5^4}{c_2c_3^3c_4}\,,\\
0=&\del_\theta a_5-\del_t a_4+\del_x a_3\,.
\end{split}
\eqlabel{maxwel}
\end{equation}

\subsection{$SO(6)$ symmetric black holes}
The static black hole solution of \eqref{eoms} and \eqref{maxwel} with
$SO(6)$ symmetry takes the form:
\begin{equation}
\begin{split}
&a_3=a_4=0\,,\qquad a_5=\frac{L^4\sin^3 x}{\cos^5 x}\,,\\
&c_1=\frac{L\sqrt{a(x)}}{\cos x}\,,\qquad c_2=\frac{L}{\sqrt{a(x)}\cos x}\,,
\qquad c_3=L \tan x\,,\qquad c_4=c_5=L\,,\\
&a(x)=1 -\frac{\mu \cos^4 x}{\sin^2 x}\,,
\qquad  \mu=\frac{r_+^2}{L^2} \left(
\frac{r_+^2}{L^2} +1\right)\,,\qquad x\in \left[0,\frac \pi2\right]\,,
\end{split}
\eqlabel{ads}
\end{equation}
where 
$L$ is the $S^5$ radius, and $r_+$ is the ``black hole size''
as measured by the $S^3$ radius at the Schwarzschild horizon.

\subsection{$SO(5)$ invariant linearized fluctuations}\label{lisone}
Now we consider $\ell=1$ fluctuations on $S^5$ about \eqref{ads}.
To this end we assume, to linear order in $\e$:
\begin{equation}
\begin{split}
&a_3=-\e L^4  A_3(x)\sin\theta e^{-i\w t}\,,\qquad a_4=-i\e L^4 A_4(x)\sin\theta e^{-i \w t}\,,\\
&a_5=\frac{L^4\sin^3 x}{\cos^5 x}\biggl(1+\e A_5(x)\cos\theta e^{-i\w t}\biggr)\,,
\qquad c_1=\frac{L\sqrt{a(x)}}{\cos x}\left(1+\e f_1(x)\cos\theta e^{-i\w t}\right)\,,\\
&c_2=\frac{L}{\sqrt{a(x)}\cos x}\left(1+\e f_2(x)\cos\theta e^{-i\w t}\right)\,,\qquad 
 c_3=L \tan x\left(1+\e f_3(x)\cos\theta e^{-i\w t}\right)\,,\\
&c_4=L\left(1+\e f_4(x)\cos\theta e^{-i\w t}\right)\,,\qquad c_5=L\left(1+\e f_5(x)\cos\theta e^{-i\w t}\right)\,.
\end{split}
\eqlabel{anzatsfl}
\end{equation}
It is easy to see that the five-form flux thorough the $S^5$ is unchanged to $\calo(\e)$.

Next, substituting \eqref{anzatsfl} into \eqref{eoms} and \eqref{maxwel} we find at $\calo(\e)$:
\nxt from $G_{tt}=0$,
\begin{equation}
\begin{split}
&0=f_1''+\frac{8 \tan^2 x A_5}{(\mu \cos^4 x-\sin^2 x)}
+\frac{\w^2\sin^4 x  (4 f_5+f_2+f_4+3 f_3)}{(\mu \cos^4 x-\sin^2 x)^2}-
\frac{3 \tan^2 x (f_1+8 f_3)}{\mu \cos^4 x-\sin^2 x}\\
&-\frac{\tan x (2 \mu \cos^4 x-\cos^2 x+4) f_1'}{\mu \cos^4 x-\sin^2 x}
+\frac{(\mu \cos^4 x+\sin^4 x)(f_2'-3 f_3'-f_4'-4 f_5')}{\cos x (\mu \cos^4 x-\sin^2 x) \sin x} \,,
\end{split}
\eqlabel{11}
\end{equation}
\nxt from $G_{xx}=0$,
\begin{equation}
\begin{split}
&0=f_1''+4 f_5''+f_4''+3 f_3''
+\frac{8 \tan^2 x A_5}{\mu \cos^4 x-\sin^2 x}+\frac{f_2 \w^2 \sin^4 x}
{(\mu \cos^4 x-\sin^2 x)^2}\\
&+\frac{1}{\sin x \cos x (\mu \cos^4 x-\sin^2 x)}
\biggl((\cos^4 x (-5+2 \cos^2 x) \mu-\sin^4 x) f_1'\\
&+(\sin^2 x (-\cos^2 x+4)-2 \mu \cos^4 x) f_2'
+(3 \cos^4 x (2 \cos^2 x-1) \mu\\
&-3 \sin^2 x (\cos^2 x+1)) f_3'+(\cos^4 x (2 \cos^2 x-3) \mu+\sin^4 x) f_4'
+(4 \cos^4 x (2 \cos^2 x-3) \mu\\
&+4 \sin^4 x) f_5')
+\frac{\tan^2 x (5 f_2-8 f_1-24 f_3)}{\mu \cos^4 x-\sin^2 x}\,,
\end{split}
\eqlabel{22}
\end{equation}
\nxt from $G_{\Om_3\Om_3}=0$,
\begin{equation}
\begin{split}
0=&f_3''+\frac{8 \tan^2 x A_5}{\mu \cos^4 x-\sin^2 x}
+\frac{f_3\w^2 \sin^4 x }{(\mu \cos^4 x-\sin^2 x)^2}
+\frac{f_1'-f_2'+f_4'+4 f_5'}{\cos x \sin x}\\
&+\frac{2 (\mu \cos^6 x+\mu \cos^4 x+3 \cos^2 x-3) f_3'}
{\cos x (\mu \cos^4 x-\sin^2 x) \sin x}
+\frac{1}{(\mu \cos^4 x-\sin^2 x) \cos^2 x}(4 f_2 \cos^2 x\\
&-8 \sin^2 x f_1+(15 \cos^2 x-19) f_3)\,,
\end{split}
\eqlabel{33}
\end{equation}
\nxt from $G_{\theta\theta}=0$,
\begin{equation}
\begin{split}
0=&f_4''+\frac{(2 \mu \cos^6 x-\mu \cos^4 x+3 \cos^2 x-3) f_4'}
{\cos x (\mu \cos^4 x-\sin^2 x) \sin x}-\frac{8 \tan^2 x A_5}{\mu \cos^4 x-\sin^2 x}
\\&+\frac{f_4 \w^2 \sin^4 x }{(\mu \cos^4 x-\sin^2 x)^2}
+\frac{3 \tan^2 x (3 f_2+9 f_3+4 f_5+3 f_1-4 f_4)}{\mu \cos^4 x-\sin^2 x}\,,
\end{split}
\eqlabel{44}
\end{equation}
\nxt from $G_{\Om_4\Om_4}=0$ we obtain two ODEs,
\begin{equation}
\begin{split}
0=&f_5''+\frac{(2 \mu \cos^6 x-\mu \cos^4 x+3 \cos^2 x-3) f_5'}
{\cos x (\mu \cos^4 x-\sin^2 x) \sin x}-\frac{8 \tan^2 x A_5}{\mu \cos^4 x-\sin^2 x}
\\
&+\frac{f_5 \w^2\sin^4 x }{(\mu \cos^4 x-\sin^2 x)^2}
+\frac{3 \tan^2 x (3 f_1+3 f_2+9 f_3-f_4+f_5)}{\mu \cos^4 x-\sin^2 x}\,,
\end{split}
\eqlabel{5050}
\end{equation}
\begin{equation}
\begin{split}
0=&f_5''+\frac{(2 \mu \cos^6 x-\mu \cos^4 x+3 \cos^2 x-3) f_5'}
{\cos x (\mu \cos^4 x-\sin^2 x) \sin x}
-\frac{8 \tan^2 x A_5}{\mu \cos^4 x-\sin^2 x}\\&
+\frac{f_5 \w^2\sin^4 x }{(\mu \cos^4 x-\sin^2 x)^2}
+\frac{3 \tan^2 x (3 f_2+9 f_3+3 f_5+3 f_1-3 f_4)}{\mu \cos^4 x-\sin^2 x}\,,
\end{split}
\eqlabel{5252}
\end{equation}
\nxt from $G_{tx}=0$,
\begin{equation}
\begin{split}
0=&\w\biggl[f_5'+\frac14 f_4'+\frac34 f_3'+\frac{1}{\cos x (\mu \cos^4 x-\sin^2 x) \sin x}
+\biggl(\frac34 (-\mu \cos^4 x+\sin^2 x) f_2\\
&+\left(\frac32 \mu \cos^4 x-\frac34 \cos^2 x \sin^2 x\right) f_3
+(\mu \cos^4 x+\sin^4 x) \left(f_5+\frac 14 f_4\right)\biggr)
\biggr]\,,
\end{split}
\eqlabel{12}
\end{equation}
\nxt from $G_{t\theta}=0$  we obtain two ODEs,
\begin{equation}
\begin{split}
0=&A_4+ \frac{\w \sin x^5 (f_2+3 f_3+4 f_5)}{8\cos^3 x (\mu \cos^4 x-\sin^2 x)}\,,
\end{split}
\eqlabel{160}
\end{equation}
\begin{equation}
\begin{split}
0=&A_4+ \frac{\w \sin x^5 (f_2+3 f_3+8 f_5-4 f_4)}{8\cos^3 x (\mu \cos^4 x-\sin^2 x)}\,,
\end{split}
\eqlabel{162}
\end{equation}
\nxt from $G_{x\theta}=0$  we obtain two ODEs,
\begin{equation}
\begin{split}
0=&f_5'+\frac14 f_1'+\frac34 f_3'-\frac{2 \cos^3 x A_3}{(\mu \cos^4 x-\sin^2 x) \sin x}
+\frac{3f_3}{4\cos x \sin x}\\
&-\frac{(\mu \cos^4 x+\cos^4 x-2 \cos^2 x+1) f_1}{4\sin x \cos x (\mu \cos^4 x-\sin^2 x)}
-\frac{(2 \mu \cos^4 x-\cos^4 x+5 \cos^2 x-4) f_2}{4(\mu \cos^4 x-\sin^2 x) \cos x \sin x}\,,
\end{split}
\eqlabel{260}
\end{equation}
\begin{equation}
\begin{split}
0=&f_5'+\frac18 f_1'+\frac38 f_3'-\frac12 f_4'
-\frac{\cos^3 x A_3}{(\mu \cos^4 x-\sin^2 x) \sin x}+ \frac{3f_3}{8\cos x \sin x}\\
&- \frac{(\mu \cos^4 x+\cos^4 x-2 \cos^2 x+1) f_1}{8\sin x \cos x (\mu \cos^4 x-\sin^2 x)}
-\frac{(2 \mu \cos^4 x-\cos^4 x+5 \cos^2 x-4) f_2}{8(\mu \cos^4 x-\sin^2 x) \cos x \sin x}\,,
\end{split}
\eqlabel{262}
\end{equation}
\nxt from \eqref{maxwel},
\begin{equation}
\begin{split}
0=&A_5'+f_4'-f_1'-3 f_3'-f_2'+4 f_5'-\frac{5 \cos^3 x A_3}{(\mu \cos^4 x-\sin^2 x) \sin x}\,,
\end{split}
\eqlabel{bi1}
\end{equation}
\begin{equation}
\begin{split}
0=&A_4-\frac{\w \sin^5 x (-f_4+f_2-A_5+3 f_3-4 f_5+f_1)}{5\cos^3 x (\mu \cos^4 x-\sin^2 x)}\,,
\end{split}
\eqlabel{bi2}
\end{equation}
\begin{equation}
\begin{split}
0=& A_4'+\frac{\sin^4 x A_3 \w}{(\mu \cos^4 x-\sin^2 x)^2}
+\frac{(2 \mu \cos^6 x-7 \mu \cos^4 x+3 \sin^2 x) A_4}{(\mu \cos^4 x-\sin^2 x) \sin x \cos x}\,,
\end{split}
\eqlabel{bi3}
\end{equation}
\begin{equation}
\begin{split}
0=& A_3'+\frac{\sin^3 x A_5}{\cos^5 x}-A_4 \w\,.
\end{split}
\eqlabel{bi4}
\end{equation}
Note that in \eqref{12} we have kept the factor of $\w$ --- this will be important in what follows.

\section{Hubeny-Rangamani zero mode}\label{hr}
To search for the zero mode, ~\cite{Hubeny:2002xn} set $\w=0$ and obtained
the solution of the resulting equations. It is easy to verify that \eqref{11}-\eqref{5252} and 
\eqref{160}-\eqref{bi4} are solved with 
\begin{equation}
\begin{split}
&A_3=A_4=A_5=f_4=f_5=0\,,\\
&f_3=-\frac 13 f_1-\frac 13 f_2\,,\\
&f_1=-\frac{\tan x (\mu \cos^4 x-\sin^2 x) f_2'}{2 \cos^2 x \mu-\sin^2 x}
-\frac{(3 \mu \cos^4 x-\cos^4 x+6 \cos^2 x-5) f_2}{\cos^2 x (2 \cos^2 x \mu-\sin^2 x)}\,,\\
&0=f_2''+\frac{f_2'}{\cos x \sin x (2 \cos^2 x \mu-\sin^2 x) (\mu \cos^4 x-\sin^2 x)}
\biggl(2 \mu (1+2\mu)\cos^8 x\\
&+(6 \mu^2-7\mu-2) \cos^6 x+(27 \mu+11) \cos^4 x-(22\mu+16) \cos^2 x+7\biggr)
\\&
+\frac{((34 \mu+7) \cos^4 x-(54\mu+14) \cos^2 x +7) f_2}{\cos^2 x (2 \cos^2 x \mu-\sin^2 x) (\mu \cos^4 x-\sin^2 x)}\,.
\end{split}
\eqlabel{hr1}
\end{equation}
It is important to realize that \eqref{12} is consistent only because of the overall factor 
$\w$. That is, note the combination in $[ \cdots ]$ of \eqref{12} is inconsistent with the second order 
equation for $f_2$ in \eqref{hr1}.

For an explicit comparison one can
introduce $r=L\tan x$ and show that the second order equation for $f_2$ in \eqref{hr1} becomes identical 
to equation (4.8) in~\cite{Hubeny:2002xn} ($\chi(r)\equiv f_2(r)$) with $\ell=1$.  As a result, the zero frequency 
normalizable mode of \eqref{hr1} is precisely the one identified in HR.  To contrast with the discussion 
in section \ref{gl} below,  we proceed with analysis of \eqref{hr1}. We find it convenient to introduce
\begin{equation}
\begin{split}
&y=\frac{r_+}{L\tan x} = \frac{\r_+}{\tan x}\,,\qquad \r_+=\frac{r_+}{L}\,,\qquad y\in [0,1]\,,\\
&f_2(y)=y^7 F_2(y)\,,
\end{split}
\eqlabel{defy}
\end{equation} 
where $y\to 0_+$ is the AdS boundary and $z\equiv 1-y\to 0_+$ is the location of the Schwarzschild horizon.
In terms of the $y$ coordinate we now have,
\begin{equation}
\begin{split}
&0=F_2''+\frac{(11 y^2-29 y^4+22 y^6+\r_+^2(44  y^6-29 y^4-6 y^2+7)+(22  y^6-6 y^2)\r_+^4) F_2'}{
y (y^2-1) (\r_+^2+(\r_+^2+1) y^2) (2 (\r_+^2+1) y^2-1)}\\
&+\frac{((42 y^4 +16) \r_+^4+\r_+^2(84  y^4-55  y^2+16 )+42 y^4-55 y^2+21) F_2}
{(\r_+^2+(\r_+^2+1) y^2) (2 (\r_+^2+1) y^2-1) (y^2-1)}\,.
\end{split}
\eqlabel{reducedHR}
\end{equation}
The UV indices of \eqref{reducedHR} are determined assuming $F_2\sim y^n$ as $y\to 0$:
\begin{equation}
y^{n-2}\ n (n+6) =0\,.
\eqlabel{uvindex}
\end{equation} 
Naively, the normalizable mode, $n=0$, corresponds to a dimension-7 operator expectation value of the boundary SYM and
the mode with $n=-6$ is non-normalizable. This is not the case however: from \eqref{hr1} notice that if 
$f_2\sim y^n=y^7$ as $y\to 0$, $f_1 \sim y^n \cos(x)^{-2}\sim y^{n-2}=y^5$. The gauge invariant fluctuation 
would be a linear combination of $f_1$ and $f_2$ modes,  and thus would have a fall-off $\propto y^5$ corresponding to 
a massive $\ell=1$ Kaluza-Klein graviton \cite{ds}.

The appropriate boundary expansion in the UV takes form:
\begin{equation}
F_2=1-\frac{16 \r_+^2+16 \r_+^4+21}{16 \r_+^2}\ y^2 +\calo(y^4)\,,
\eqlabel{ubhr}
\end{equation}
where without loss of generality we normalized $F_2(y=0)=1$. Near the horizon we require a smooth solution, 
thus 
\begin{equation}
F_2= f_{2,h}^0\left(1+\frac{29\r_+^2+8}{4(2\r_+^2+1)}\ z+\calo(z^2)\right)\,.
\eqlabel{irhr}
\end{equation}
We use a shooting method to numerically connect the UV \eqref{ubhr} and the IR \eqref{irhr} asymptotics of 
\eqref{reducedHR}, 
tuning the two parameters $\r_+$ and $f_{2,h}^0$.  We find
\begin{equation}
f_{2,h}^0 \simeq 0.068027\,,\qquad \r_+\simeq 0.440234\,,
\eqlabel{hrres}
\end{equation}
a result first obtained in \cite{Hubeny:2002xn}.

\section{$\ell=1$ quasinormal modes}\label{gl}

We now perform the standard quasinormal analysis of $\ell=1$ mode: we analyze \eqref{11}-\eqref{bi4} 
for general $\w$. Consistency of \eqref{5050} and \eqref{5252} implies that 
\begin{equation}
f_5=f_4\,.
\eqlabel{consta}
\end{equation}
From \eqref{160} (or \eqref{162})\,
\begin{equation}
A_4=-\frac{\omega \sin^5 x (f_2+3 f_3+4 f_4)}{8\cos^3 x (\mu \cos^4 x-\sin^2 x)}\,.
\eqlabel{constb}
\end{equation}
Next, from \eqref{bi2},
\begin{equation}
A_5=f_1-\frac 52 f_4 +\frac{13}{8} f_2+\frac{39}{8} f_3\,.
\eqlabel{constc}
\end{equation}
With a simple algebra, the remaining equations  
are reduced to a system of coupled first-order equations for $f_1$,  $f_2$,  $f_3$,  $f_4$,  $A_3$ :
\begin{equation}
\begin{split}
&0=f_1'+\frac{\sin^5 x (3 f_3+5 f_4) \cos x\ \w^2}{8(\mu \cos^4 x-\sin^2 x)^2}
-\frac{5 A_3 \cos^3 x}{\sin x (\mu \cos^4 x-\sin^2 x)}\\
&-\frac{1}{8\sin x \cos x (\mu \cos^4 x-\sin^2 x)^2} \biggl(5 \mu \cos^4 x (\mu \cos^4 x+\cos^2 x-1) f_1+(3 \sin^6 x\\
&+\sin^2 x \cos^4 x (3 \cos^2 x+5) \mu-8 \mu^2 \cos^8 x) f_2+(3 \sin^6 x (\cos^2 x-5)\\
&-6 \sin^2 x \cos^4 x (\cos^2 x+3) \mu+21 \mu^2 \cos^8 x) f_3+(5 \sin^6 x (\cos^2 x-4)\\
&-5 \sin^2 x \cos^4 x (\cos^2 x+6) \mu+30 \mu^2 \cos^8 x) f_4\biggr)\,,
\end{split}
\eqlabel{fin1}
\end{equation}
\begin{equation}
\begin{split}
&0=f_2'-\frac{5 A_3 \cos^3 x}{\sin x (\mu \cos^4 x+\cos^2 x-1)}+\frac{\sin^5 x (3 f_3+5 f_4) \cos x\ \w^2}{8(\mu \cos^4 x+\cos^2 x-1)^2}
\\&+\frac{1}{8\sin x \cos x (\mu \cos^4 x+\cos^2 x-1)^2} \biggl((-8 \sin^6 x-\sin^2 x \cos^4 x (8 \cos^2 x-5) \mu\\
&+3 \mu^2 \cos^8 x) f_1+(-\sin^4 x (5 \cos^2 x-29)+\sin^2 x \cos^4 x (5 \cos^2 x-53) \mu\\
&+24 \mu^2 \cos^8 x) f_2+(-9+30 \cos^4 x+3 \cos^8 x-24 \cos^6 x+6 \sin^2 x \cos^4 x (\cos^2 x\\
&+11) \mu
-45 \mu^2 \cos^8 x) f_3+(-5 \sin^6 x (\cos^2 x-4)+5 \sin^2 x \cos^4 x (\cos^2 x+6) \mu\\
&-30 \mu^2 \cos^8 x) f_4\biggr)\,,
\end{split}
\eqlabel{fin2}
\end{equation}
\begin{equation}
\begin{split}
&0=f_3'-\frac{5 A_3 \cos^3 x}{\sin x (\mu \cos^4 x+\cos^2 x-1)}- \frac{5\sin^5 x (3 f_3+5 f_4) \cos x\ \w^2}{24(\mu \cos^4 x+\cos^2 x-1)^2}
\\&-\frac{1}{24\sin x \cos x (\mu \cos^4 x+\cos^2 x-1)^2} \biggl(
(-40 \sin^6 x-5 \sin^2 x \cos^4 x (8 \cos^2 x-5) \mu\\&+15 \mu^2 \cos^8 x) 
f_1+(-\sin^4 x (25 \cos^2 x-49)+\sin^2 x \cos^4 x (25 \cos^2 x-73) \mu\\
&+24 \mu^2 \cos^8 x) f_2
+(3 \sin^4 x (5 \cos^4 x+2 \cos^2 x-15)-6 \sin^2 x \cos^4 x (11 \cos^2 x-23) \mu\\
&-33 \mu^2 \cos^8 x) f_3
+(-5 \sin^6 x (5 \cos^2 x+12)-15 \sin^2 x \cos^4 x (9 \cos^2 x-10) \mu\\
&+10 \mu^2 \cos^8 x) f_4\biggr)\,,
\end{split}
\eqlabel{fin3}
\end{equation}
\begin{equation}
\begin{split}
&0=f_4'+\frac{3 A_3 \cos^3 x}{\sin x (\mu \cos^4 x+\cos^2 x-1)}+\frac{\sin^5 x (3 f_3+5 f_4) \cos x\ \w^2}{8(\mu \cos^4 x+\cos^2 x-1)^2}
\\&+\frac{1}{8\sin x \cos x (\mu \cos^4 x+\cos^2 x-1)^2} \biggl((-8 \sin^6 x-\sin^2 x \cos^4 x (8 \cos^2 x-5) \mu\\
&+3 \mu^2 \cos^8 x) f_1+(5 \sin^6 x-5 \sin^4 x \cos^4 x \mu) f_2+(-3 \sin^6 x (\cos^2 x+3)\\
&+18 \sin^4 x \cos^4 x \mu+3 \mu^2 \cos^8 x) f_3+(-5 \sin^6 x (\cos^2 x+4)-5 \sin^2 x \cos^4 x (7 \cos^2 x\\
&-6) \mu+10 \mu^2 \cos^8 x) f_4\biggr)\,,
\end{split}
\eqlabel{fin4}
\end{equation}
\begin{equation}
\begin{split}
&0=A_3'+\frac{\sin^5 x (f_2+3 f_3+4 f_4) \w^2}{8\cos^3 x (\mu \cos^4 x+\cos^2 x-1)}+\frac{\sin^3 x (8 f_1+39 f_3-20 f_4+13 f_2)}{8\cos^5 x}\,.
\end{split}
\eqlabel{fin5}
\end{equation}
We can further algebraically eliminate $f_1$, $f_2$ and $A_3$ from \eqref{fin1}-\eqref{fin5}. Using the radial coordinate $y$ as in 
\eqref{defy}, we obtain a system of coupled ODEs for $f_3$ and $f_4$: 
\begin{equation}
\begin{split}
&0=f_4''-\r_+^2\left(\r_+^2 y^4+\r_+^2+y^4\right)^{-1} f_3''+(8 \r_+^2 y^4+7 \r_+^2+8 y^4-3 y^2) \r_+^2y^{-1} (-y^2+y^4-\r_+^2\\
&+\r_+^2 y^4)^{-1} 
(\r_+^2 y^4+\r_+^2+y^4)^{-1} f_3'+\frac13 (3 \r_+^4 y^8+57 \r_+^4 y^4+59 \r_+^4+6 \r_+^2 y^8+3 \r_+^2 y^6\\
&+57 \r_+^2 y^4-7 \r_+^2 y^2+3 y^8+3 y^6)y^{-1} (-y^2+y^4-\r_+^2+\r_+^2 y^4)^{-1} (\r_+^2 y^4+\r_+^2+y^4)^{-1} f_4'\\
&-(21 \r_+^4 y^8+52 \r_+^4 y^4-35 \r_+^4+42 \r_+^2 y^8-14 \r_+^2 y^6+52 \r_+^2 y^4-47 \r_+^2 y^2+y^2 \w^2 \r_+^2+21 y^8\\
&-14 y^6) \r_+^2(-y^2+y^4-\r_+^2+\r_+^2 y^4)^{-2} (\r_+^2 y^4+\r_+^2+y^4)^{-1} y^{-2} f_3+\frac13 
(30 \r_+^4 y^8-35 \r_+^4 y^4\\
&-185 \r_+^4+60 \r_+^2 y^8+3 \r_+^2 y^6 \w^2-65 \r_+^2 y^6-35 \r_+^2 y^4-125 \r_+^2 y^2+3 y^2 \w^2 \r_+^2+30 y^8\\
&+3 y^6 \w^2-65 y^6) \r_+^2 f_4(-y^2+y^4-\r_+^2+\r_+^2 y^4)^{-2} (\r_+^2 y^4+\r_+^2+y^4)^{-1} y^{-2}\,,
\end{split}
\eqlabel{eq1}
\end{equation}
\begin{equation}
\begin{split}
&0=f_3'''-\frac13 (6 \r_+^4 y^8+27 \r_+^4 y^4+50 \r_+^4+12 \r_+^2 y^8-9 \r_+^2 y^6+27 \r_+^2 y^4-7 \r_+^2 y^2+6 y^8\\
&-9 y^6)y^{-1} (-y^6+y^8-\r_+^2 y^2-\r_+^2 y^6+2 \r_+^2 y^8-\r_+^4
+\r_+^4 y^8)^{-1} f_3''+\frac13 (-12 \r_+^6 y^{12}\\
&+330 \r_+^6 y^8+538 \r_+^6 y^4+167 \r_+^6
-36 \r_+^4 y^{12}+9 y^{10} \r_+^4+660 \r_+^4 y^8+3 y^6 \w^2 \r_+^4-188 \r_+^4 y^6
\\&+538 \r_+^4 y^4-286 \r_+^4 y^2+3 \r_+^4 y^2 \w^2-36 \r_+^2 y^{12}+18 \r_+^2 y^{10}+339 \r_+^2 y^8+3 \r_+^2 y^6 \w^2\\
&-188 \r_+^2 y^6+3 \r_+^2 y^4-12 y^{12}
+9 y^{10}+9 y^8)y^{-2} (-2 y^6-2 \r_+^2 y^4+2 \r_+^2 y^8+\r_+^4 y^8+y^8\\
&-2 \r_+^2 y^6-2 \r_+^4 y^4+2 \r_+^2 y^2+\r_+^4+y^4)^{-1} (\r_+^2 y^4+\r_+^2+y^4)^{-1} f_3'
+\frac59 (318 \r_+^6 y^8+759 \r_+^6 y^4\\
&+470 \r_+^6+24 y^{10} \r_+^4+636 \r_+^4 y^8-48 \r_+^4 y^6+759 \r_+^4 y^4-128 \r_+^4 y^2+48 \r_+^2 y^{10}+318 \r_+^2 y^8\\
&-48 \r_+^2 y^6
-4 \r_+^2 y^4+24 y^{10})y^{-2} (\r_+^2 y^4+\r_+^2+y^4)^{-1} (-y^2+y^4-\r_+^2+\r_+^2 y^4)^2)^{-1} f_4'\\
&-\frac13 (24 y^6 \w^2 \r_+^4-12 y^8 \w^2 \r_+^2-10 y^4 \w^2 \r_+^4+38 y^2 \w^2 \r_+^6-1550 \r_+^6 y^6-1550 \r_+^4 y^6\\
&-1365 \r_+^6 y^{10}+470 \r_+^4 y^4+3132 \r_+^4 y^{12}-2730 y^{10} \r_+^4+1044 \r_+^2 y^{12}-1365 \r_+^2 y^{10}\\
&+1304 \r_+^6 y^4+3132 \r_+^6 y^{12}+1304 y^4 \r_+^8+1044 y^{12} \r_+^8+36 y^{14} \r_+^4+12 y^{14} \r_+^6+36 y^{14} \r_+^2\\
&+3198 \r_+^6 y^8+2135 \r_+^4 y^8+15 \r_+^6 y^{10} \w^2+12 y^{14}-1165 \r_+^8-1262 \r_+^6 y^2+536 \r_+^2 y^8\\
&+30 y^{10} \w^2 \r_+^4-12 y^8 \w^2 \r_+^4+15 \r_+^2 \w^2 y^{10}+24 y^6 \w^2 \r_+^6+1599 \r_+^8 y^8)(-y^2+y^4-\r_+^2\\
&+\r_+^2 y^4)^{-3}
 (\r_+^2 y^4+\r_+^2+y^4)^{-1} y^{-3} f_3
+\frac59 (39 y^6 \w^2 \r_+^4-6 y^8 \w^2 \r_+^2-6 y^4 \w^2 \r_+^4+21 y^2 \w^2 \r_+^6\\
&-655 \r_+^6 y^6-655 \r_+^4 y^6-660 \r_+^6 y^{10}+250 \r_+^4 y^4+513 \r_+^4 y^{12}-1320 y^{10} \r_+^4+171 \r_+^2 y^{12}\\
&-660 \r_+^2 y^{10}-1214 \r_+^6 y^4+513 \r_+^6 y^{12}-1214 y^4 \r_+^8+171 y^{12} \r_+^8\\
&-36 y^{14} \r_+^4-12 y^{14} \r_+^6-36 y^{14} \r_+^2-1038 \r_+^6 y^8-245 \r_+^4 y^8+18 \r_+^6 y^{10} \w^2-12 y^{14}\\
&-1220 \r_+^8-403 \r_+^6 y^2+274 \r_+^2 y^8+36 y^{10} \w^2 \r_+^4-6 y^8 \w^2 \r_+^4+18 \r_+^2 \w^2 y^{10}+39 y^6 \w^2 \r_+^6\\
&-519 \r_+^8 y^8)(-y^2+y^4-\r_+^2+\r_+^2 y^4)^{-3} (\r_+^2 y^4+\r_+^2+y^4)^{-1} y^{-3} f_4\,.
\end{split}
\eqlabel{eq2}
\end{equation}

To determine the UV indices of \eqref{eq1} and \eqref{eq2} we substitute $f_3\sim s_3 y^n\,, f_4\sim s_4 y^n$. Requiring 
nontrivial solution as $y\to 0_+$ determines the 5 indices of  \eqref{eq1} and \eqref{eq2} in the UV:
\begin{equation}
(n-5)(n+5)(n-9)(n+1)^2=0\,.
\eqlabel{uvfull}
\end{equation} 
Note that there are two independent normalizable modes in the UV: one associated with the expectation 
value of the dimension-5 operator, and the other one is associated with the dimension-9 operator.

To determine the IR indices of \eqref{eq1} and \eqref{eq2} we substitute $f_3\sim q_3 (1-y)^n\,, f_4\sim q_4 (1-y)^n$. Requiring 
nontrivial solution as $z=(1-y)\to 0_+$ determines the 5 indices of  \eqref{eq1} and \eqref{eq2} in the IR:
\begin{equation}
\begin{split}
&0=(1-2n)\left(\left(\frac{\w}{4\pi T}\right)^2+n^2\right)\left(\left(\frac{\w}{4\pi T}\right)^2+(n-1)^2\right)\,,\\
&T=\frac{2\r_+^2+1}{2\pi\r_+}\,,
\end{split}
\eqlabel{itfull}
\end{equation} 
where $T$ is the black hole temperature. Thus, near the horizon we have (we assume $\Re(w)\ge 0$):
\nxt 2 incoming modes:
\begin{equation}
n=-i\ \frac{\w}{4\pi T}\,,\qquad  n=1-i\ \frac{\w}{4\pi T}\,,
\eqlabel{incoming}
\end{equation}
\nxt 2 outgoing modes:
\begin{equation}
n=+i\ \frac{\w}{4\pi T}\,,\qquad  n=1+i\ \frac{\w}{4\pi T}\,,
\eqlabel{outgoing}
\end{equation}
\nxt and 1 ``localized'' mode:
\begin{equation}
n=\frac 12\,.
\eqlabel{loc}
\end{equation}

To determine the spectrum of the quasinormal modes we require that $f_3$ and $f_4$ radial wavefunctions are normalizable 
near the AdS boundary and are either incoming or localized at the horizon. 
Assuming that\footnote{This is validated by our finding of solutions from the followup computations.} 
the quasinormal mode is purely dissipative (or unstable),  
\begin{equation}
\Re(\w)=0\,,\qquad \Im(\w)=-i g\,,
\eqlabel{diss}
\end{equation}
we have the following asymptotic expansion in the UV, 
\begin{equation}
\begin{split}
&f_{3}=y^5 (1-y)^{-i \w/(4\pi T)}\biggl(
1-\frac12 \r_+ g\ \frac{y}{2 \r_+^2+1}+\biggl(-\frac{83}{56} \r_+^2-\frac{83}{224 \r_+^2}
-\frac{83}{56}+\biggl(-\frac14 \r_+\\
&-\frac12 \r_+^3\biggr) g
+\biggl(\frac{5}{14} \r_+^2+\frac{13}{224 \r_+^2}+\frac{13}{56}\biggr) g^2\biggr) 
\left(\frac{y}{2 \r_+^2+1}\right)^2
+\biggl(\biggl(\frac{83}{448 \r_+}+\frac{25}{336} \r_+^3-\frac23 \r_+^5\\
&+\frac{193}{336} \r_+\biggr) g+\biggl(\frac18 \r_+^2+\frac14 \r_+^4\biggr) g^2+\biggl(-\frac{23}{168} \r_+^3
-\frac{13}{448 \r_+}-\frac{13}{112} \r_+\biggr) g^3\biggr) \left(\frac{y}{2 \r_+^2+1}\right)^3\\
&+f_{3,4} y^4+\calo(y^5)\,,
\biggr)
\end{split}
\eqlabel{f3uv}
\end{equation}
\begin{equation}
\begin{split}
&f_{4}=y^5 (1-y)^{-i \w/(4\pi T)}\biggl(
-\frac17+\frac{1}{14} \r_+ g\ \frac{y}{2 \r_+^2+1}
+\biggl(\frac{29}{56} \r_+^2+\frac{29}{224 \r_+^2}+\frac{29}{56}
+\biggl(\frac{1}{14} \r_+^3\\
&+\frac{1}{28} \r_+\biggr) g
+\biggl(-\frac{1}{14} \r_+^2-\frac{3}{56}-\frac{3}{224 \r_+^2}\biggr) g^2\biggr) \left(\frac{y}{2 \r_+^2+1}\right)^2
+\biggl(\biggl(-\frac{55}{336} \r_+^3
+\frac{2}{21} \r_+^5\\
&-\frac{29}{448 \r_+}-\frac{79}{336} \r_+\biggr) g+\biggl(-\frac{1}{28} \r_+^4
-\frac{1}{56} \r_+^2\biggr) g^2
+\biggl(\frac{3}{112} \r_++\frac{3}{448 \r_+}+\frac{5}{168} \r_+^3\biggr) g^3\biggr) \\
&\times\left(\frac{y}{2 \r_+^2+1}\right)^3
+\biggl(-\frac{1448}{875} \r_+^2-\frac{492}{125} \r_+^4
+\frac{13}{500 \r_+^4}+\frac{114}{875 \r_+^2}-\frac{3264}{875} \r_+^6-\frac{24}{5} f_{3,4} \r_+^2
\\&-\frac{72}{5} \r_+^4 f_{3,4}-\frac{96}{5} \r_+^6 
f_{3,4}-\frac{48}{5} \r_+^8 f_{3,4}-\frac35 f_{3,4}-\frac{1088}{875} \r_+^8-\frac{66}{875}
+\biggl(\frac{23}{280} \r_++\frac{13}{560 \r_+}\\
&-\frac12 \r_+^5-\frac{16}{35} \r_+^7
-\frac{9}{140} \r_+^3\biggr) g
+\biggl(\frac{22}{105} \r_+^6-\frac{2567}{4200} \r_+^2-\frac{27}{1400 \r_+^4}
-\frac{611}{4200} \r_+^4-\frac{2657}{5600}\\
&-\frac{27}{175 \r_+^2}\biggr) g^2
+\biggl(-\frac{3}{560 \r_+}-\frac{11}{140} \r_+^3-\frac{9}{280} \r_+-\frac{1}{14} \r_+^5\biggr) g^3
+\biggl(\frac{3}{7000 \r_+^4}+\frac{197}{10500} \r_+^4\\
&+\frac{171}{7000} \r_+^2+\frac{3}{875 \r_+^2}+\frac{363}{28000}
 \biggr)g^4\biggr) \left(\frac{y}{2 \r_+^2+1}\right)^4+\calo(y^5)
\biggr)\,,
\end{split}
\eqlabel{f4uv}
\end{equation}
where without loss of the generality we have fixed the coefficient of the leading asymptote of $f_3$ to be 1.
In the IR we have,
\begin{equation}
\begin{split}
&f_3=(1-z)^5 z^{-i\w/(4\pi T)}\biggl(f_{3,b,0}+f_{3,b,1} z+\calo(z^2)\biggr)
+(1-z)^5 z^{1/2} f_{3,a,0}\biggl(1+\calo(z)\biggr)\,,
\end{split}
\eqlabel{f3ir}
\end{equation}
\begin{equation}
\begin{split}
&f_4=(1-z)^5 z^{-i\w/(4\pi T)}\biggl(-\frac 35 f_{3,b,0}+
 \frac35 \biggl(
\biggl((6\r_+^4+5\r_+^2)g^2+(-72\r_+^3-68\r_+^5-19\r_+)g\\
&+20+520\r_+^6+600\r_+^4+210\r_+^2\biggr)f_{3,b,0}
+\biggl((-5\r_+^3-10\r_+^5)g+5\r_+^2+20\r_+^6\\
&+20\r_+^4\biggr)f_{3,b,1}
\biggr)(13\r_+^2+4)^{-1}(\r_+g-1-2\r_+^2)^{-1}(2\r_+^2+1)^{-1}\ z
+\calo(z^2)\biggr)
\\&+(1-z)^5 z^{1/2} f_{3,a,0}\biggl(\frac{\r_+^2}{2\r_+^2+1}+\calo(z)\biggr)\,.
\end{split}
\eqlabel{f4ir}
\end{equation}
In \eqref{f3uv}-\eqref{f4ir} we presented terms only to the order necessary to demonstrate the dependence on
all the coefficients\footnote{To ensure accuracy, in the numerical analysis we have kept further terms in the asymptotic 
expansions.}:
\begin{equation}
\{g\,,\, f_{3,4}\,,\, f_{3,a,0}\,,\,  f_{3,b,0}\,,\, f_{3,b,1}\,\}\,.
\eqlabel{coeff}
\end{equation}
Notice that we have 5 integration constants \eqref{coeff}--- precisely the number 
needed to uniquely identify the solution of one second order \eqref{eq1} and one third order \eqref{eq2} equation,
given the value of $\r_+$.

\begin{figure}[t]
\begin{center}
\psfrag{x}{{$\r_+$}}
\psfrag{y}{{$-\frac{\Im\w}{2\pi T}$}}
\includegraphics[width=3.0in]{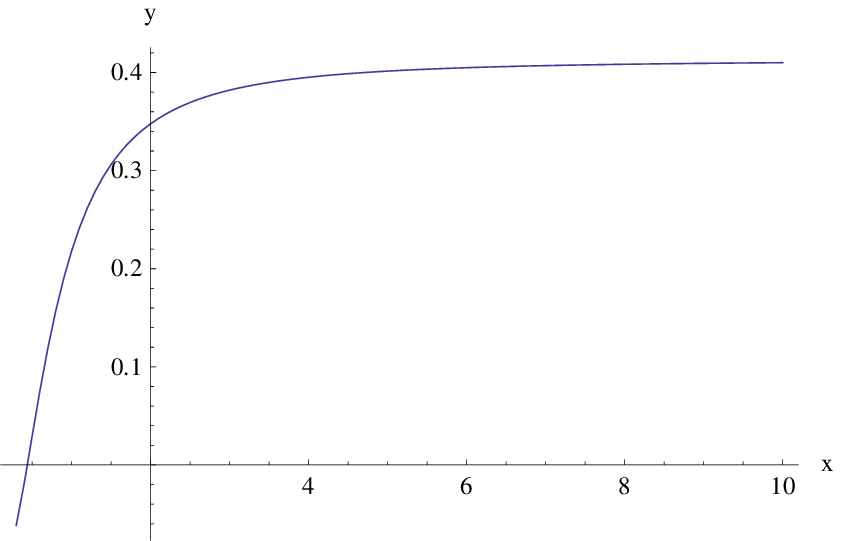}
\includegraphics[width=3.0in]{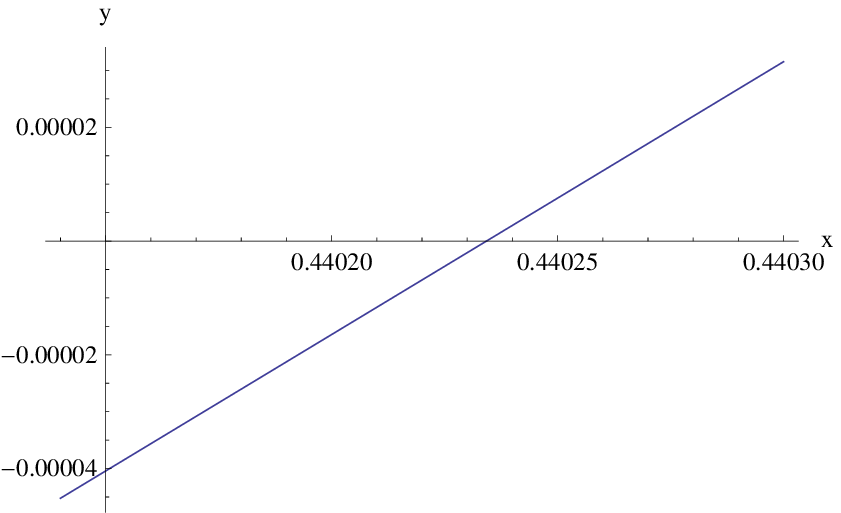}
\end{center}
  \caption{ The dependence of the $g=-\Im(\w)$ as a function of 
$\r_+=\frac{r_+}{L}$ for $\ell=1$ fluctuations of $SO(6)$ symmetric black holes in 
$AdS_5\times S^5$. Black holes with $g<0$ are unstable with respect to condensation of 
these fluctuations.} \label{figure1}
\end{figure}

We can use the shooting method of \cite{Aharony:2007vg} to compute $g(\r_+)$. 
The results of the computations are presented in figure \ref{figure1}.
The left panel represents the dependence of the $g=-\Im(\w)$ as a function of 
$\r_+=\frac{r_+}{L}$. Black holes with $g>0$ are stable, while the black holes with 
$g<0$ are unstable with respect to the condensation of $\ell=1$ fluctuations 
discussed in this section. The right panel  shows the transition region. 
We find that 
\begin{equation}
\Im(w(\r_+))=0\qquad \Longleftrightarrow\qquad \r_+\simeq 0.440234\,,
\eqlabel{final}
\end{equation}
which, to six significant digits, coincides with the location of the HR zero mode \eqref{hrres}.

\section{Higher-$\ell$ quasinormal modes}\label{higher}

In the previous section we demonstrated that the $\ell=1$ quasinormal mode becomes unstable 
for  sufficiently small black hole, see \eqref{final}.  Here, we argue that 
higher-$\ell$ quasinormal modes do not affect such conclusion. Further, we reveal that 
the leading  instability is indeed due to  the $\ell=1$ mode. 

Rather than generalizing the analysis of section \ref{gl} for $\ell>1$ quasinormal modes, we present an
alternative computation of the spectrum. We choose the `transverse-traceless' gauge (TT) for the fluctuations:
\begin{equation}
\begin{split}
&c_1=\frac{L\sqrt{a(x)}}{\cos x}\left(1+\epsilon f_1(x) Y_{\ell}(\theta) e^{-i\w t}\right)\,,\qquad c_2=\frac{L}{\sqrt{a(x)}\cos x}\left(1+\epsilon f_2(x) Y_{\ell}(\theta) e^{-i\w t}\right)\,,\\
&c_3=L\tan x\left(1-\frac 13\epsilon (f_1(x)+f_2(x)) Y_{\ell}(\theta) e^{-i\w t}\right)\,,\qquad g_{tx}=\epsilon f_{tx}(x) Y_{\ell}(\theta) e^{-i\w t}\,,
\end{split}
\eqlabel{ttgauge}
\end{equation} 
where $Y_\ell$ are the $S^5$-spherical harmonics,
\begin{equation}
\Delta_{S^5} Y_{\ell}\equiv -s\ Y_\ell=-\ell(\ell+4) Y_\ell \,.
\eqlabel{yldef}
\end{equation}

\begin{figure}[t]
\begin{center}
\psfrag{x}{{$\r_+$}}
\psfrag{y}{{$-\frac{\Im\w}{2\pi T}$}}
\psfrag{q}{{$s$}}
\psfrag{w}{{$\r_+^2$}}
\includegraphics[width=3.0in]{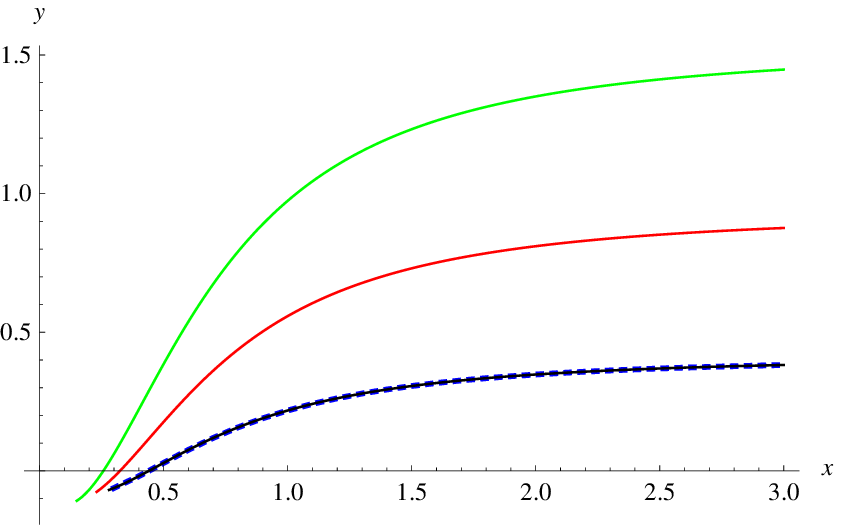}
\includegraphics[width=3.0in]{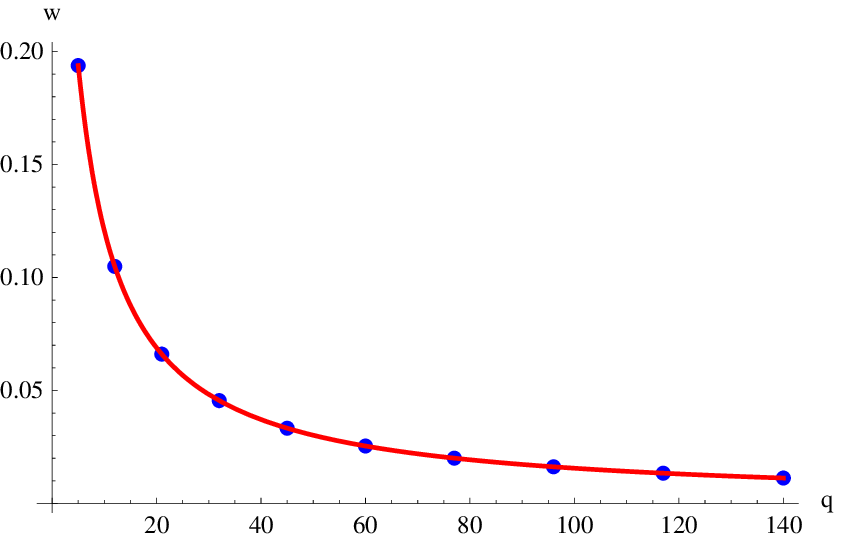}
\end{center}
  \caption{ Left panel: the dependence of the $g=-\Im(\w)$ as a function of 
$\r_+=\frac{r_+}{L}$ for $\ell=\{1,2,3\}$ (solid lines: black/red/green) 
fluctuations of $SO(6)$ symmetric black holes in 
$AdS_5\times S^5$ in transverse-traceless gauge. 
The dashed blue line represents $\ell=1$ results of 
figure \ref{figure1} obtained in a 'diagonal' gauge \eqref{diffeo}.
Black holes with $g<0$ are unstable with respect to condensation of 
these fluctuations --- the leading instability is due to $\ell=1$ mode.
Right panel: the critical values of $\r_+^2$ (blue dots) 
at which higher $s=\ell(\ell+4)$ quasinormal modes become unstable. 
The solid red line is the fit to the data, see \eqref{fit}. 
} \label{figure2}
\end{figure}

The benefit of the TT gauge \eqref{ttgauge} is that it is consistent to 
set all the other fluctuations of the metric and the 5-form to 
zero\footnote{We are indebted to Jorge Santos for pointing this to us. We also explicitly verified this.}.
Furthermore, $f_1$ and $f_2$ can be algebraically expressed in 
terms of $f_{tx}$ and its derivative $f_{tx}'$. The latter function satisfies
the following second order ODE:
\begin{equation}
\begin{split}
&0=f_{tx}''+ (y (\r_+^4 y^8+2 \r_+^2 y^8+6 \r_+^4 y^4+6 \r_+^2 y^4-3 \r_+^4+y^8+\r_+^4 y^4 s+\r_+^2 y^4 s-4 \r_+^2 y^2\\
&-\r_+^2 y^2 s-\r_+^4 s
+\omega^2 \r_+^2 y^2) (\r_+^2 y^2+\r_+^2+y^2) (\r_+^2+y^2) (y^2-1))^{-1}(\r_+^2 (y^2-1) (\r_+^2 y^2\\
&+\r_+^2+y^2) (7 \r_+^4 y^4+\r_+^4+3 \r_+^2 y^6
+7 \r_+^2 y^4+2 \r_+^2 y^2+3 y^6+y^4) s+\r_+^2 y^2 (9 \r_+^4 y^4+3 \r_+^4\\
&+5 \r_+^2 y^6+9 \r_+^2 y^4+4 \r_+^2 y^2+5 y^6+y^4) \omega^2+(-1-3 \r_+^4
-\r_+^6-3 \r_+^2) y^{14}+(17 \r_+^2+3 \r_+^8\\
&+7+16 \r_+^4+9 \r_+^6) y^{12}+(34 \r_+^2+34 \r_+^6+68 \r_+^4) y^{10}+(51 \r_+^8+49 \r_+^4-2 \r_+^2
+102 \r_+^6) y^8\\
&+(-9 \r_+^6-9 \r_+^4-4 \r_+^2) y^6+(-3 \r_+^8-3 \r_+^6-13 \r_+^4) y^4-12 \r_+^6 y^2-3 \r_+^8) f_{tx}'+
((y^2-1)^2 \\
&(\r_+^4 y^8+2 \r_+^2 y^8+6 \r_+^4 y^4+6 \r_+^2 y^4-3 \r_+^4+y^8+\r_+^4 y^4 s+\r_+^2 y^4 s
-4 \r_+^2 y^2-\r_+^2 y^2 s\\
&-\r_+^4 s+\omega^2 \r_+^2 y^2) (\r_+^2 y^2+\r_+^2+y^2)^2 (\r_+^2+y^2)^2 y^2)^{-1} ((\r_+^8+4 \r_+^2
+6 \r_+^4+1+4 \r_+^6) y^{20}\\
&+(-42 \r_+^2-60 \r_+^4-24 \r_+^8-12-6 \r_+^{10}-48 \r_+^6) y^{18}+(4 \r_+^{10}-128 \r_+^6-6 \r_+^2 s-116 \r_+^4
\\&+\r_+^{12}-18 \r_+^4 s-38 \r_+^8-18 \r_+^6 s-14 \r_+^2-6 \r_+^8 s+15) y^{16}+(-168 \r_+^8-58 \r_+^6+164 \r_+^4\\
&+110 \r_+^2-56 \r_+^{10}+16 \r_+^4 s
+14 \r_+^2 s-36 \r_+^8 s-22 \r_+^6 s-12 \r_+^{10} s+2 \omega^2 \r_+^6+4 \omega^2 \r_+^4\\
&+2 \omega^2 \r_+^2) y^{14}+(309 \r_+^8
+498 \r_+^6+207 \r_+^4-30 \r_+^2+72 \r_+^{10}+24 \r_+^{12}+31 \r_+^4 s+2 \r_+^{12} s\\
&-11 \r_+^2 s+48 \r_+^8 s+86 \r_+^6 s+6 \r_+^{10} s
+16 \omega^2 \r_+^6+14 \omega^2 \r_+^4+8 \omega^2 \r_+^8+6 \omega^2 \r_+^2+\r_+^4 s^2\\
&+\r_+^8 s^2+2 \r_+^6 s^2) y^{12}+(280 \r_+^8
+52 \r_+^6-88 \r_+^4+12 \r_+^2+140 \r_+^{10}-46 \r_+^4 s+3 \r_+^2 s\\
&+52 \r_+^8 s-20 \r_+^6 s+26 \r_+^{10} s+42 \omega^2 \r_+^6+28 \omega^2 \r_+^4
+14 \omega^2 \r_+^{10}+28 \omega^2 \r_+^8-3 \omega^2 \r_+^2-2 \r_+^4 s^2\\
&+2 \r_+^{10} s^2+4 \r_+^8 s^2+2 \omega^2 \r_+^6 s
+2 \omega^2 \r_+^4 s) y^{10}+(-52 \r_+^8-58 \r_+^6+51 \r_+^4+12 \r_+^{10}
+6 \r_+^{12}\\
&+17 \r_+^4 s+\r_+^{12} s^2-2 \r_+^{12} s-67 \r_+^8 s-65 \r_+^6 s-4 \r_+^{10} s+34 \omega^2 \r_+^6-16 \omega^2 \r_+^4
+34 \omega^2 \r_+^8\\
&+\r_+^4 s^2+2 \r_+^{10} s^2-5 \r_+^8 s^2-6 \r_+^6 s^2+\omega^4 \r_+^4+4 \omega^2 \r_+^8 s+4 \omega^2 \r_+^6 s-2 \omega^2 \r_+^4 s) y^8+(24 \r_+^8\\
&+90 \r_+^6+24 \r_+^{10}-36 \r_+^8 s+39 \r_+^6 s-36 \r_+^{10} s-29 \omega^2 \r_+^6+12 \omega^2 \r_+^{10}+12 \omega^2 \r_+^8\\
&-6 \r_+^{10} s^2-6 \r_+^8 s^2+4 \r_+^6 s^2
+2 \omega^4 \r_+^6+2 \omega^2 \r_+^{10} s+2 \omega^2 \r_+^8 s-6 \omega^2 \r_+^6 s) y^6+(84 \r_+^8+24 \r_+^{10}\\
&+24 \r_+^{12}
-2 \r_+^{12} s^2-6 \r_+^{12} s+45 \r_+^8 s-6 \r_+^{10} s-22 \omega^2 \r_+^8-2 \r_+^{10} s^2+6 \r_+^8 s^2+\omega^4 \r_+^8
\\&-6 \omega^2 \r_+^8 s) y^4+(42 \r_+^{10}+26 \r_+^{10} s-6 \omega^2 \r_+^{10}+4 \r_+^{10} s^2-2 \omega^2 \r_+^{10} s) y^2
+(s+3)^2 \r_+^{12}) f_{tx}\,.
\end{split}
\eqlabel{ftx}
\end{equation}
where we used the radial coordinate $y$ as in \eqref{defy}.
Further introducing 
\begin{equation}
f_{tx}=y^{\ell+3} (1-y)^{-i\omega/(4\pi T)-1}F_{tx}\,, 
\eqlabel{strip}
\end{equation}
the quasinormal spectra for different $\ell$ are determined solving \eqref{ftx} with the boundary conditions
\begin{equation}
\lim_{y\to 0_+} F_{tx}={\rm finite}\,,\qquad \lim_{y\to 1_-} F_{tx}=1\,.
\eqlabel{ftxbc}
\end{equation}  
The quasinormal spectra for $\ell=\{1,2,3\}$ are presented in figure 
\ref{figure2} (left panel) as solid black/red/green lines. 
The  dashed blue line represents $\ell=1$
spectrum in a 'diagonal gauge'  \eqref{diffeo}.  
The $\ell=1$ mode is the first one to go unstable as the 
size of the black hole is decreased.

Both the equation \eqref{ftx} and the asymptotic expansions resulting 
from the boundary conditions \eqref{ftxbc} are smooth in the 
limit $\omega\to 0$. Thus, we can directly determine the critical values 
$\r_+^2$ when the higher harmonics $s=\ell(\ell+4)$ become unstable. 
The results of such analysis are shown as blue dots in figure 2 
(right panel), and are collected in the table:
\begin{center}
  \begin{tabular}{ | c | c | c | c | c | c  | }
 \hline
 $\ell$ & 1 & 2 & 3& 4& 5\\   \hline
 $\r_+^2$ &0.440234   &0.323890	
 &0.257042 	 & 0.213323	&
0.182412  \\  
\hline\hline 
 $\ell$ & 6 & 7 & 8& 9& 10\\   \hline
 $\r_+^2$ &0.159369 	 &0.141516 	 
&0.127273 	 & 0.115641 
&  0.105962 	 \\ 
\hline
  \end{tabular}
\end{center}
The solid red line in the right panel of figure \ref{figure2} is the fit to the data:
\begin{equation}
\r_+^2\bigg|_{fit}=\frac{1.60892}{s}-\frac{5.35936}{s^2}
+\frac{16.1219}{s^3}-\frac{26.6116}{s^4}+\calo(s^{-5})\,.
\eqlabel{fit}
\end{equation} 
In leading $\frac 1s$ dependence of $\r_+^2$ in \eqref{fit} 
can be independently verified 
as follows. Introducing 
\begin{equation}
\r_+^2=\frac{q_0}{s}+\calo(s^{-2})\,,
\eqlabel{defq0}
\end{equation}
the $\omega=0$ equation \eqref{ftx} to leading order in $\frac 1s$ takes form:
\begin{equation}
\begin{split}
&0=f_{tx}''+(y (y^2-1) (y^6+q_0 y^2-q_0))^{-1}(-y^8+7 y^6+3 q_0 y^4-2 q_0 y^2-q_0) 
f_{tx}'\\
&+((y^6+q_0 y^2-q_0) (y^2-1)^2 y^4)^{-1}(y^{12}-12 y^{10}
-6 q_0 y^8+15 y^8+14 q_0 y^6+q_0^2 y^4\\
&-11 q_0 y^4
-2 q_0^2 y^2+3 q_0 y^2+q_0^2) f_{tx}\,.
\end{split}
\eqlabel{largel}
\end{equation}
Because of the explicit linear dependence on $\ell$ in the $y\to 0_+$
limit, $f_{tx}\propto y^{\ell}$ \eqref{strip},  the $\ell\to \infty$
limit of the $y\to 0_+$ boundary condition is modified to an essential 
singularity in \eqref{largel}: 
\begin{equation}
f_{tx}\propto \exp\left(\pm \frac{\sqrt{q_0}}{y}\right)\,.
\end{equation}
Introducing 
\begin{equation}
f_{tx}=  \exp\left(- \frac{\sqrt{q_0}}{y}\right) (1-y)^{-1} F_{tx}\,,
\eqlabel{largeftx}
\end{equation}
the boundary conditions for $F_{tx}$ are kept as in \eqref{ftxbc}. 
Solving the resulting ODE we find
\begin{equation}
q_0=1.61015\,,
\eqlabel{q0res}
\end{equation}
in excellent agreement with the leading $\frac 1s$ factor in the 
fit \eqref{fit}. 

Although the near-boundary fall-off of the $f_{xt}$ is $\propto y^{\ell+3}$ \eqref{ftxbc}, 
 $f_1\propto y^{\ell+4}$, which implies that the dual SYM operator is of dimension $\Delta=\ell+4$.
Of course, the UV indices of the radial profiles of the quasinormal modes provide 
only a suggestive identification of the unstable modes with the operators of the dual 
$\caln=4$ SYM. A precise identification requires a complete holographic discussion
along the lines of \cite{Skenderis:2006uy}. Such analysis are   
beyond the scope of this paper\footnote{See \cite{Dias:2015pda} for a relevant discussion.}. 
Note however that the instability of small black holes is clearly due to a massive Kaluza-Klein 
graviton --- this is particularly clear in the TT gauge where the 5-form modes are not exited.
A simple analysis show that a massless minimally coupled scalar (mimicking the 10-dimensional 
graviton) with an angular momentum $\ell$ along the $S^5$ has a radial profile with the fall-off 
$\propto y^{\ell+4}$ near the boundary, corresponding to a dimension $\Delta=\ell+4$ operator of the dual SYM. 
This observation is the primary reason of our identification of the instability on the CFT side.

\section{Conclusion}\label{con}

In this paper we computed the spectra of quasinormal modes, up to $l=10$, of $SO(6)$ 
symmetric black holes 
in $AdS_5\times S^5$. We showed that these black holes are in principle subject to a GL-type
 instability once they become sufficiently small. 
The location of the onset of GL instability coincides with the black hole size when 
the $\ell=1$ HR zero mode becomes normalizable. (Modes with larger $l$ require smaller black holes
to reach the instability onset).
Having identified this  possibility, one could speculate
the behavior of the instability by analogy to black strings  subject to the GL instability~\cite{Gregory:1993vy}. 
There, the separatrix between stable and unstable black strings is identified precisely by a zero mode.
Furthermore, in the unstable regime, as shown in~\cite{Lehner:2010pn,Lehner:2011wc} a self-similar behavior ensues indicating
that  within a finite affine-time a horizon pinch-off reveals a naked singularity. If an analog behavior were to
take place in $AdS_5\times S^5$, it would imply that naked singularities can naturally develop in gravitational theories. 
And, through holography, it would open a venue to understand such development from the perspective of 
non-gravitational quantum field theories. Whether this behavior is indeed realized awaits future inspection.


\section*{Acknowledgments}
We would like to thank Oscar Dias, Veronika Hubeny, Mukund Rangamani and Jorge Santos for valuable correspondence
and comments on the draft.
Oscar Dias and Jorge Santos independently computed the growth rates of the 
GL unstable mode discussed in this paper.  
Research at Perimeter
Institute is supported by the Government of Canada through Industry
Canada and by the Province of Ontario through the Ministry of
Research \& Innovation. This work was further supported by
NSERC through the Discovery Grants program (AB and LL) and CIFAR (LL).

\end{document}